\documentstyle[11pt,psfig,fullpage,citesort]{article}

\newcommand{\ket}[1]{|#1\rangle}
\newcommand{\bra}[1]{\langle#1|}

\def\m@th{\mathsurround=0pt }
\def\leftrightarrowfill{$\m@th \mathord\leftarrow \mkern-6mu
 \cleaders\hbox{$\mkern-2mu \mathord- \mkern-2mu$}\hfill
 \mkern-6mu \mathord\rightarrow$}
\def\overleftrightarrow#1{\vbox{\ialign{##\crcr
     \leftrightarrowfill\crcr\noalign{\kern-1pt\nointerlineskip}
     $\hfil\displaystyle{#1}\hfil$\crcr}}}
\def\VEV#1{\langle#1\rangle}
\def\phdag{{\phantom{\dagger}}}
\begin{document}
\renewcommand{\thefootnote}{\fnsymbol{footnote}}
\begin{titlepage}
\begin{flushright}
UFIFT-HEP-00-6\\
hep-th/0003301
\end{flushright}

\vskip 1.5cm

\begin{center}
\begin{Large}
{\bf Spontaneous Symmetry Breaking \\ 
at Infinite Momentum without $P^+$ Zero-Modes}
\end{Large}

\vskip 2.cm

{\large Joel S. Rozowsky\footnote{E-mail  address: rozowsky@phys.ufl.edu} and Charles B. Thorn\footnote{E-mail  address: thorn@phys.ufl.edu}}

\vskip 0.5cm

{\it Institute for Fundamental Theory\\
Department of Physics, University of Florida,
Gainesville, FL 32611}

(\today)

\vskip .5cm
\end{center}

\begin{abstract}
\noindent The nonrelativistic interpretation of quantum field theory 
achieved by quantization in an infinite momentum frame is spoiled
by the inclusion of a mode of the field carrying $p^+=0$. We
therefore explore the viability of doing without such a mode in
the context of spontaneous symmetry breaking (SSB), where its
presence would seem to be most needed. We show that the physics of
SSB in scalar quantum field theory in $1+1$ space-time dimensions
is accurately described without a zero-mode.
\end{abstract}
\end{titlepage}

The infinite momentum frame provides a vehicle for casting
any relativistic quantum mechanical system in terms of the
(nonrelativistic) quantum dynamics of Heisenberg and 
Schr\"odinger inspired by the classical dynamics of Galileo and Newton 
\cite{diracfront}. This
Newtonian view of quantum field theories \cite{weinbergfront,susskindgal}
might arguably be dismissed as a mere curiosity, since those theories have
several satisfactory manifestly relativistic formulations. But
the corresponding view of string theory \cite{goddardgrt} remains one of the
best hopes for a truly fundamental description of string that does
not rely on perturbation theory \cite{thornmosc}. Thus it behooves us to
probe the viability of the Newtonian view of quantum field theory
\cite{thorn}, since the
latter might well be merely a low energy effective theory for string.

In quantum field theory the best way to achieve the nonrelativistic
description is to employ light front coordinates \cite{diracfront}
$x^\pm=(x^0\pm x^3)/\sqrt2$, choosing $x^+$ as time, 
and referring the $x^-$ coordinate to its conjugate momentum
labelled by $p^+>0$. Then $p^+$ assumes the role of a
variable Newtonian mass. A typical quantum field, for instance
a real scalar field $\phi$, has the expansion
\begin{eqnarray}
\phi({\bf x},x^-,x^+)&=&\int_0^\infty {dp^+\over\sqrt{4\pi p^+}}
\left(a({\bf x},p^+)e^{-ix^-p^+} +a^\dagger({\bf x},p^+)e^{+ix^-p^+}
\right),
\end{eqnarray}
where the quantum nature of the field is fixed by imposing
\begin{eqnarray}
\left[a({\bf x},p^+),a^\dagger({\bf y},q^+)\right]&=& 
\,\delta({\bf x}-{\bf y})\delta(p^+-q^+) \nonumber\\
\left[a({\bf x},p^+),a({\bf y},q^+)\right] &=& \,0.
\end{eqnarray}
Discretizing $p^+=lm$, $l=1,2,\ldots$, sharpens the Newtonian
interpretation, because then each value of $l$ labels a different
species of particle with Newtonian mass $lm$\cite{thorn}. 
However, much debate
and controversy in the discretized light-cone quantization (DLCQ)
literature (for a review see \cite{brodskyppreport}) has centered on the 
possible necessity of including a field mode carrying 
$p^+=0$~\cite{zeromodes}. 
It has been suggested that without such a mode it would be difficult
if not impossible to describe
such a commonplace field theoretic phenomenon as spontaneous
symmetry breaking \cite{zeromodessb}. If a zero-mode were really
necessary, we would have no good Newtonian interpretation for it, and
the field theory would not be adequately described by Newtonian dynamics.

While conceding that the inclusion of a zero-mode is a valid
field theoretic option, we argue in this letter that it is
not necessary, even to describe spontaneous symmetry
breaking. First of all, the physics of
condensation associated with SSB does not require a fundamental
zero-mode. Just consider the Cooper pairs of BCS superconductivity,
which most definitely carry Newtonian mass. Similarly, in
the infinite momentum frame there is no compelling reason to
require that a condensate carry zero $p^+$. It is only necessary
that in the infinite volume limit, local physics cannot extract
$p^+$ from or deposit $p^+$ into the condensate.

In this letter, we shall give a detailed analysis
of the simplest field theory that exhibits SSB, namely a
real scalar field in 1 space dimension. At the outset, our
quantum field will have no zero-mode. We shall show that, in spite of this,
the physics of SSB is completely and accurately described in
our model. 
 
\begin{figure}[ht]
\centerline{\psfig{file=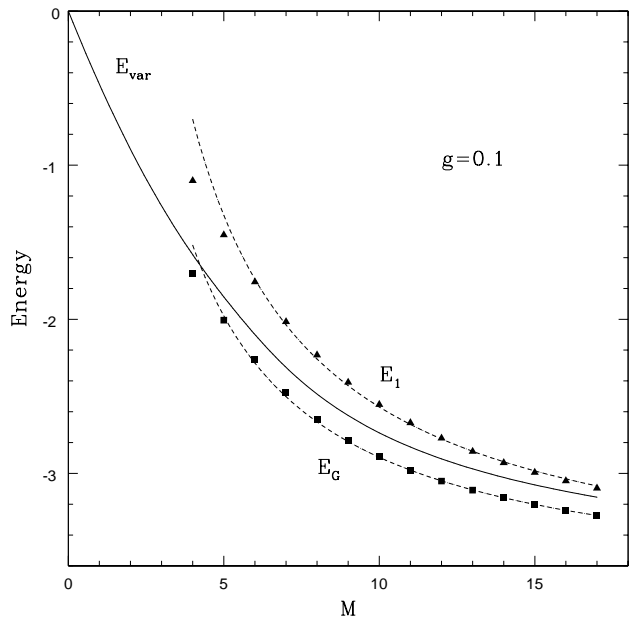,width=3.5in}}
\caption[]{Energy as a function of $M$ of the ground state ($E_G$) 
and the first excited state ($E_1$) compared to the variational calculation
(solid curve) -- for $g=0.1$. The dashed curves are fits to the
eigenvalues in the range $10\leq M\leq 17$ as described later in the text.}
\label{fig1}
\end{figure}

\begin{figure}[ht]
\centerline{\psfig{file=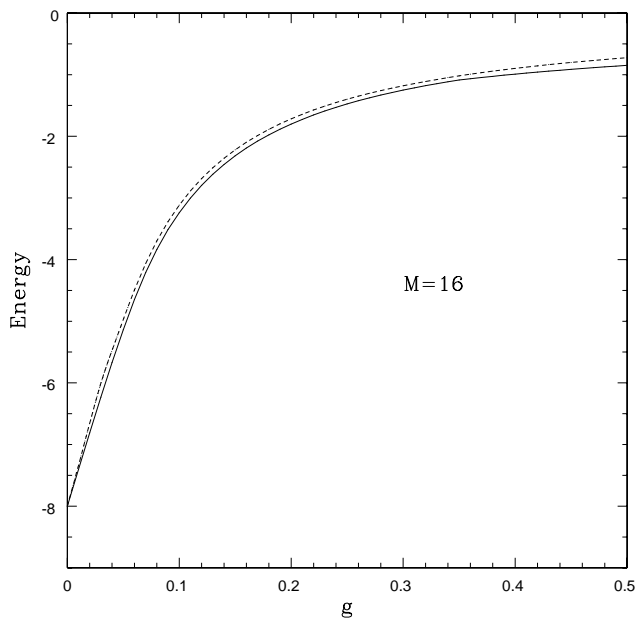,width=3.5in}}
\caption[]{The energy dependence on $g$ (for $M=16$) of 
the variational calculation (dashed curve) compared to the 
numerical calculation for the lowest eigenstate (solid curve).}
\label{fig2}
\end{figure}

The light-cone Hamiltonian is $P^-$,\, the density of which we choose to be
\begin{eqnarray}
{\cal H}=-{\mu^2\over2}:\!\phi^2\!:+{\lambda\over24}:\!\phi^4\!:.
\end{eqnarray}
With discrete $p^+=lm$ (equivalently periodic boundary conditions in
$x^-$ on the interval $-\pi/m<x^-<\pi/m$), the field has the expansion
\begin{eqnarray}
\phi=\sum_{l=1}^\infty{1\over\sqrt{4\pi l}}\left[a^\phdag_le^{il\theta}
+a^\dagger_le^{-il\theta}\right],
\end{eqnarray}
where we have defined the angle $\theta=-mx^-$. The quantum conditions
are then simply $[a_j,a^\dagger_l]=\delta_{jl}$, $[a_j,a_l]=0$. Note
the complete absence of a zero-mode. It is convenient to also define
a rescaled Hamiltonian, $h\equiv mH/\mu^2$, 
\begin{eqnarray}
h&=&-\sum_{l>0}{a^\dagger_la^{\phdag}_l\over2l}+
{g\over4}\sum_{l_1+l_2>l_3>0}{a^\dagger_{l_1+l_2-l_3}a^\dagger_{l_3}
a^\phdag_{l_1}a^\phdag_{l_2}\over\sqrt{l_1l_2l_3(l_1+l_2-l_3)}} \\
&&\hskip 0cm
+{g\over6}\sum_{l_1,l_2,l_3>0}{{a^\dagger_{l_1+l_2+l_3}a^\phdag_{l_3}
a^\phdag_{l_1}a^\phdag_{l_2}}+{a^\dagger_{l_1}a^\dagger_{l_2}
a^\dagger_{l_3}a^\phdag_{l_1+l_2+l_3}}\over\sqrt{l_1l_2l_3(l_1+l_2+l_3)}},
\nonumber
\end{eqnarray}
where $g\equiv\lambda/8\pi\mu^2$. The dynamical system has now been
completely specified. The negative quadratic term is designed to
drive the instability towards spontaneous symmetry breaking.
The Hamiltonian possesses a parity symmetry under $\phi\to-\phi$,
and also conserves discrete total $P^+=Mm$. A brute force way to analyze 
the dynamics would be to look for energy eigenstates with definite
$P^+$, that is fixed $M$. The state space in this subspace has
dimension $p(M)$, the number of unordered partitions of the integer $M$.
As long as $M$ is not too large ($p(M)$ increases
exponentially with $\sqrt{M}$), the Hamiltonian can be numerically
diagonalized with the aid of a computer (results of our analysis for
$M\leq17$ can be seen in Fig.~\ref{fig1} and Fig.~\ref{fig2}).

However, actual symmetry breaking can occur only in the infinite
volume limit. This is taken by letting $m\to0$ and $M\to\infty$
keeping $P^+=Mm$ fixed. Symmetry breaking occurs when the lowest
energy level is degenerate, possessing two states of opposite
parity. But at finite volume, tunneling
between these states always lifts the degeneracy. Thus all one can
hope to find in the brute force numerical method is a gradual
approach to degeneracy as $M$ is increased. Our initial studies
in this direction were equivocal: we could not deal with sufficiently
large $M$ to definitively reveal such a trend. 

Since the model is expected to show symmetry breaking for arbitrarily weak
coupling, we should be able to confirm this analytically.
Since the weak coupling limit is semi-classical, it is natural 
to apply the variational principle, choosing as a trial the coherent
state\footnote{A coherent state involving only zero-modes was used in
an earlier attempt to understand the vacuum of this field
theory~\cite{harindranathv}.}
$\ket{\alpha}=e^{\sum\alpha_l a^\dagger_l}\ket{0}
e^{-\sum\alpha_l\alpha^*_l/2}$. The energy functional is then
\begin{eqnarray}
\bra{\alpha}h\ket{\alpha}={1\over2\pi}\int_{-\pi}^\pi d\theta\left[
-{1\over4}f^2+{g\over24}f^4\right],
\end{eqnarray}
with
\begin{eqnarray}
f(\theta)=\sum_{l=1}^\infty{1\over\sqrt{l}}\left[\alpha_le^{il\theta}
+\alpha^*_le^{-il\theta}\right].
\end{eqnarray}
The minimum occurs when $f^2(\theta)=3/g$. Since there is no
zero-mode, the simplest solution of this condition 
is given by
\begin{eqnarray}
f(\theta)=\cases{+\sqrt{3/g}& for $0\leq|\theta|<\pi/2$\cr
-\sqrt{3/g}& for $\pi/2<|\theta|<\pi$.\cr}
\end{eqnarray}
Then the $\alpha$'s are determined to be
\begin{eqnarray}
\alpha_{2n}=0, \qquad \alpha_{2n+1}={2\over\pi}
\sqrt{3\over g}{(-)^n\over\sqrt{2n+1}}.
\end{eqnarray}
Because of the discontinuities, the expectation value of 
$P^+/m=\sum_l la^\dagger_l a^\phdag_l$ in this
trial state is infinite: $\sum l|\alpha_l|^2=(12/\pi^2 g)\sum_n1$. 

There are two ways to extend this variational approach to the situation
of finite $M$. One is to work with  the projection of the
state $\ket{\alpha}$ to the subspace with definite $M$:
$\ket{\alpha, M}=P_M\ket{\alpha}$.  This is easy enough for specific
$M$ values but difficult to do for general $M$.  
A more tractable approach is to do a constrained variation: 
minimizing $H$ subject to the
constraint $\VEV{P^+}=mM$, with $M$ the number of $P^+$ units. 
This can be done by adding a Lagrange multiplier term $\beta(P^+/m-M)$
to $h$ and minimizing the expectation of the resulting 
operator. $\beta$ can then be adjusted so that the
constraint is satisfied. Instead of $\bra{\alpha}h\ket{\alpha}$, 
we now minimize
\begin{eqnarray}
\bra{\alpha}h_\beta\ket{\alpha}=
{1\over2\pi}\int_{-\pi}^\pi \hskip -.3cm d\theta\left[
\beta\left({{f^\prime}^2\over2}\!-\!M\!\right)\!-\!{1\over4}f^2\!+\!{g\over24}f^4\right],
\end{eqnarray}
again with the understanding that $f(\theta)$ 
has no zero-mode and has period $2\pi$. Denoting the maximum of $f$ by
$f_0$, the simplest solution of all these conditions is 
\begin{eqnarray}
f(\theta)=f_0\ {\rm sn}\left({2\theta+\pi
\over\pi}K,k\right),\qquad
gf_0^2={6k^2\over 1+k^2},
\end{eqnarray}
where $k$ is the modulus of the elliptic function, following the
conventions of \cite{gradshteynr}. 
Note that we have located $\theta=0$
at a maximum of the sn function. 
$K(k)$ is the quarter period of ${\rm sn}$
given by the complete elliptic integral of the first kind
\begin{eqnarray}
K=K(k)=\int_0^1 dt(1-t^2)^{-1/2}(1-k^2t^2)^{-1/2},
\end{eqnarray}
and $\beta$ is related to $k^2$ by $\beta=\pi^2/8(1+k^2)K^2$.

The constraint $\VEV{P^+}=mM$ links the last independent parameter
to $M$:
\begin{eqnarray}
M&=&{12k^2\over\pi^2g(1+k^2)^{3/2}}K\int_0^1dt\sqrt{1-t^2}
\sqrt{1-k^2t^2}\nonumber\\
&=&{4K\over\pi^2g\sqrt{1+k^2}}
\left(E-{1-k^2\over1+k^2}K\right),
\end{eqnarray}
where $E=E(k)=\int_0^1dt\sqrt{1-k^2t^2}/\sqrt{1-t^2}$ 
is the complete elliptic integral of the second kind.
We immediately see, for example, that the limit $M\to\infty$ is achieved
by taking $k^2\to1$. Indeed in this limit $K\sim(1/2)\ln(16/(1-k^2))$,
so the approach of $k^2$ to 1 is 
exponential $k^2\sim1-16e^{-M\pi^2g/\sqrt{2}}$.

We finally come to the evaluation of the expectation of $h$ in this
trial state:  
\begin{eqnarray}
\bra{\alpha}h\ket{\alpha}= -{3k^2\over2g(1+k^2)^2}+{2\over
Mg^2\pi^2(1+k^2)}\left(E-{1-k^2\over1+k^2}K\right)^2. 
\end{eqnarray}
The large $M$ behavior is easy to read off because $k^2\to1$
exponentially in $M$. Thus to any 
finite order in $1/M$ we merely set $k^2=1$
in this expression
\begin{eqnarray}
\bra{\alpha}h\ket{\alpha}\sim-{3\over8g}+{2\over g^2\pi^2}{1\over2M}
+O\left(e^{-gM\pi^2/\sqrt2}\right).
\label{varestimate}
\end{eqnarray}
The coefficient of $1/2M$ in this expression is just the square of the
 infinite volume Lorentz invariant mass of the state in units of $\mu^2$. 
 Since the state has a soliton anti-soliton pair this
mass should be twice the soliton mass, so the calculation confirms that
our setup leads to the correct soliton mass, $M_{sol}=\mu/\pi g\sqrt2=
4\sqrt{2}\mu^3/\lambda$. 

We should stress that though our variational
estimate for the energy only gives an upper bound on the true ground
state energy at general coupling, it should actually approach
the exact answer at weak coupling $g\ll1$. (Indeed, Fig.~\ref{fig2}
shows that for $M=16$ the variational estimate is quite good up to
$g=0.5$.) This is because 
our choice of trial function reduces the variational problem to
that of the classical limit, which is equivalent to weak
coupling. More precisely, we can say that the exact ground state
energy should have the large $M$ expansion
\begin{eqnarray}
E_{\rm G} \sim
-{3\over8g}(1+O(g))+{1+O(g)\over g^2\pi^2M}
+O\left({1\over M^2}\right). 
\end{eqnarray}
Note that both of the exhibited terms dominate the corrections
provided $1/g\ll M\ll 1/g^2$.
\begin{figure}[ht]
\centerline{\psfig{file=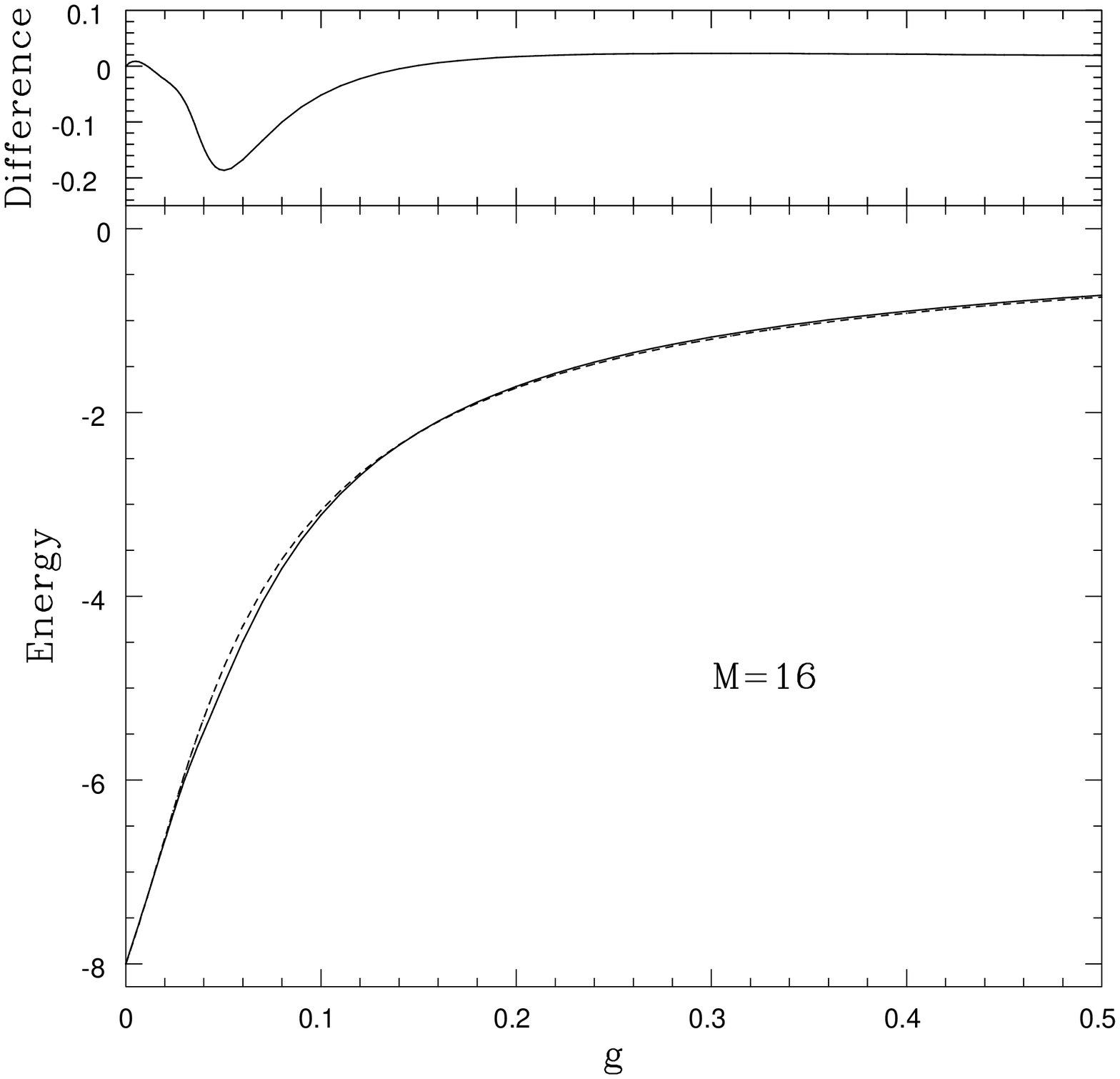,width=3.5in}}
\caption[]{Comparison of the two variational approaches for finite
$M$:  Lagrange multiplier (solid curve) and simple projection (dashed
curve) -- for $M=16$. The difference between these curves is presented
on a separate scale.}
\label{fig3}
\end{figure}
In particular, the exact and variational energies are expected to have
large $M$ limits that differ by an amount of $O(1)$ as $g\to0$.
This tendency is evident in Fig.~\ref{fig1}. More quantitatively, fits 
to the data in Fig.~\ref{fig1} for the range $10\leq M\leq 17$ 
give: $E_{\rm G}=-3.81+18.34/2M$, $E_{\rm gap}=E_1-E_{\rm
G}=6.53/2M$.  
The variational estimate of the ground state, Eq.~\ref{varestimate}, 
is $-3.75+20.26/2M$ for $g=0.1$, within 10\% of the numerical fit. 
A corresponding estimate for the gap is less direct. Fluctuations about
the trial coherent state are controlled by a $\theta$ dependent
mass squared $(gf^2(\theta)-1)\mu^2$ which approaches $2\mu^2$
as $gM\to\infty$ for almost all $\theta$. This infinite volume
value yields an estimate $2ME_{gap}\approx(4/g\pi)\approx12.73$
for $g=0.1$, nearly a factor of 2 larger than our fit. Unfortunately,
the values of $gM$ used in our fit were in the range 1 to 1.7,
for which $\int d\theta(gf^2-1)/2\pi$ varies from 1.14 to 1.49,
indicating that $12.73/2M$ is an overestimate for the gap.
In contrast, the asymptotic form of the variational energy Eq.\ref{varestimate}
is quite accurate for $gM>1$ due to the $\pi^2/\sqrt2$ in the
exponential.

The weak coupling limit also assures that the
distribution of  $M$ values in the state $\ket{\alpha}$ is sharply
peaked about its mean value $\VEV{P^+/m}$. Indeed a simple evaluation
yields at large $gM$: $\Delta M/M\approx\sqrt{3g}/\pi$ (where $\Delta
M = \sqrt{\VEV{(P^+/m)^2}-M^2}$). Thus it is in
the weak coupling limit that the constrained variational approach
is guaranteed to be equivalent to the projection onto a state of
definite $M$. Moreover, in Fig.~\ref{fig3} we see that for $M=16$ they
match quite well for $g\leq 0.5$.

We note in passing that the weak-coupling validity of the 
variational calculation also holds when $M$ stays finite (although
this is not so interesting for SSB). For example, if we examine
$g\to0$ at fixed $M$, we find that the parameter $k^2\sim 2gM/3$, so
that $\VEV{h}\sim -M/2+O(g)$, indeed tending to the minimum eigenvalue
of $h$ at $g=0$. 

We can infer the values for the $\alpha_l$ by developing $f(\theta)$
in a Fourier series:
\begin{eqnarray}
f(\theta) = {2\pi\over K}\sqrt{6/g \over 1+k^2}
\sum_{n=0}^\infty{(-)^n q^{n+1/2}\over1-q^{2n+1}}
\cos{(2n+1)\theta},
\end{eqnarray}
where $q\equiv\exp\{-\pi K(1-k^2)/K(k^2)\}$. Thus we read off
\begin{eqnarray}
\alpha_{2n}=0, \;\;
{\alpha_{2n+1}\over\sqrt{2n+1}}={\pi\over K}\sqrt{6/g \over 1+k^2}
{(-)^n q^{n+1/2}\over1-q^{2n+1}}.
\end{eqnarray}
Notice that for $k^2$ near unity, $q\sim1$, $(1-q^{2n+1})K\sim\pi^2
(2n+1)/2$, so that the $\alpha$'s revert to their step function
values, which is to be expected since this limit corresponds
to $M\to\infty$. On the other hand for $k^2$ near 0,
corresponding to weak coupling at fixed $M$, we see that
$q\to0$, so that the $\alpha_1$ mode dominates. In other words,
at weak coupling and fixed $M$ the lowest energy state is obtained by
putting all particles in the first mode, which can also be seen by
simple inspection of the $g=0$ Hamiltonian. 

Finally, we address the question of spontaneous symmetry breaking.
The trial state $\ket{\alpha}$ transforms to $\ket{-\alpha}$ under
the discrete parity transformation, and so it is not invariant
under the symmetry. Clearly the states $\ket{\pm\alpha}$
have the same variational energy. However, one can form parity eigenstates
$\ket{\pm}=C(\ket{\alpha}\pm\ket{-\alpha})$ which do not
necessarily have the same variational energy:
\begin{eqnarray}
\bra{\pm}h\ket{\pm}={\bra{\alpha}h\ket{\alpha}\pm{\rm Re}
\bra{-\alpha}h\ket{\alpha}\over1\pm{\rm Re}
\VEV{-\alpha|\alpha}}.
\end{eqnarray}
The parity eigenstates also have slightly different mean
$P^+/m$ values:
\begin{eqnarray}
\bra{\pm}{P^+\over m}\ket{\pm}
=M{1\mp{\rm Re}
\VEV{-\alpha|\alpha}\over1\pm{\rm Re}
\VEV{-\alpha|\alpha}},
\end{eqnarray}
and one must take care to compare energies at the {\it same} $\VEV{P^+}$.
However for large $M$, the $M$ dependence of $E$ is already
suppressed, and thus this subtlety can be ignored. One can 
show that for large $M$ the overlap 
$\VEV{-\alpha|\alpha}\sim(Mg\sqrt2)^{-12/\pi^2g}$.
This shows that at finite $M$ (finite volume) the states $\ket{\pm}$
are not quite degenerate, and the lower one does not break the
symmetry. However, in the infinite volume limit $M\to\infty$,
$\VEV{-\alpha|\alpha}\to0$ and degeneracy between opposite parity 
trial states is achieved, signalling SSB. Note that the splitting must
vanish faster than $1/M$, because the energy scale of the 
infinite volume theory is set by $1/M$ (recall that the 
true Hamiltonian is $(\mu^2/m)h$ and $Mm=P^+$ is fixed in the
infinite volume limit). Within the variational approximation
this puts an upper bound on the coupling for SSB, namely
$g<12/\pi^2$. Of course, the variational
solution is guaranteed to be exact only as $g\rightarrow 0$,
so that the precise value of this upper limit should be treated with
caution.  

\underline{Acknowledgments:} We thank J. Klauder and S. Shabanov for
helpful discussions. This work is supported in part by U.S. DOE grant 
DE-FG02-97ER-41029.

\end{document}